   \documentclass{aa}
   \input psfig.sty
   
\begin{document}

   \voffset -1 true cm

   \thesaurus{07.24.1; 09.07.1; 16.02.1}

   \title{Metallicity distribution of bulge planetary nebulae and
         the [O/Fe] $\times$ [Fe/H] relation}

   \author{W. J. Maciel}

   \offprints{W. J. Maciel}

   \institute{Instituto Astron\^omico e Geof\'{\i}sico, Universidade 
        de S\~ao Paulo, Av. Miguel Stefano 4200, 04301-904 S\~ao 
        Paulo SP, Brazil\\ e-mail: maciel@iagusp.usp.br}
   
   \titlerunning{Metallicity distribution of bulge PN}
   \authorrunning{W.J. Maciel}

   \date{Received date; accepted date}
   
   \maketitle

   \begin{abstract}

The O/H metallicity distribution of different samples of 
planetary nebulae in the bulge of the Milky Way
and M31 are compared. O/H abundances are converted
into [Fe/H] metallicity by the use of theoretical [O/Fe] 
$\times$ [Fe/H] relationships both for the bulge and the solar
neighbourhood. It is found that these relationships
imply an offset of [Fe/H] abundances by a factor up to
0.5 dex for bulge nebulae. Systematic errors in the O/H
abundances as suggested by some recent recombination line
work, ON cycling and statistical uncertainties are 
unable to explain the observed offset, suggesting that the 
adopted relationship for the bulge probably overestimates 
the oxygen enhancement relative to iron.

   \end{abstract}

   \section{Introduction}

Recent chemical evolution models usually predict different 
relationships between the [$\alpha$-elements/Fe] and the metallicity 
as measured by the [Fe/H] ratio for the different phases that comprise 
the Galaxy, namely the disk, bulge and halo (see for example Pagel 
\cite{pagel}). These relationships basically reflect the rate at which 
these elements are produced in different scenarios, being usually faster 
in the bulge and halo compared to the disk in the framework of
an inside-out model for Galaxy formation.

Metallicities of bulge stars are poorly known compared with disk
objects, as only limited samples of well measured stars are 
available. As a conclusion, the derived metallicity distribution and
the corresponding ratios between the $\alpha$-elements and
metalllicity are not well known. In this work,
a sample of bulge planetary nebulae (PN) with relatively accurate
abundances is used to shed some light on the [O/Fe] $\times$
[Fe/H] relationship adopted for the bulge. It will
be shown that this relation probably exagerates the
amount of oxygen produced at a given metallicity.

    \begin{figure}
    \centerline{\psfig{figure=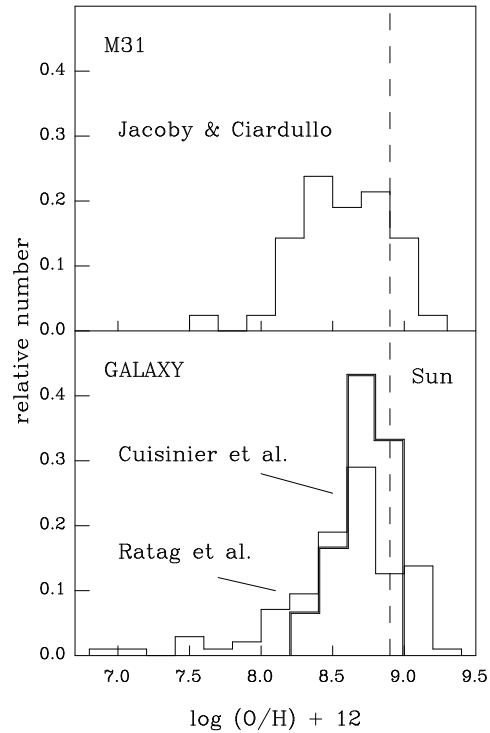,height=10.0 cm,angle=0}}
    \caption{Lower panel: metallicity distribution of bulge PN
    from Cuisinier et al. (1999, thick line) and Ratag et al. (1997,
    thin line). Top panel: The same for the PN in the bulge of M31 
   (Jacoby \&  Ciardullo 1999)}
    \label{pnoh}
    \end{figure}
   
   \section{Metallicity distribution of bulge PN}

Many bulge, or type V PN (Maciel \cite{maciel89}) are known,
but only recently have accurate abundances been obtained. 
Recent work by Cuisinier et al. (\cite{cuisinier}) and Costa \& Maciel 
(\cite{costa}) has led to He, O, N, Ar and S abundances for about 40 
bulge PN, with an uncertainty comparable to disk objects, namely up 
to 0.2 dex for O/H. The bulge O/H abundances are  generally comparable 
with those of the disk, and the O/H, Ar/H and S/H ratios can be higher 
than the disk counterparts even though very metal rich PN are missing 
in the bulge. Since underabundant nebulae are also present, these results
suggest that the bulge contains a mixed population, so that star 
formation in the bulge  spans a wide time interval. 

    \begin{figure*}
    \psfig{figure=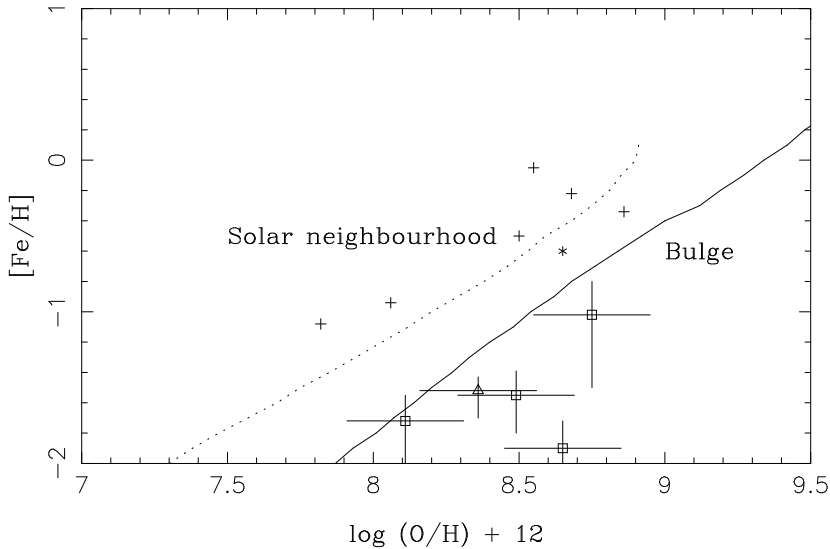,height=8.5 cm,angle=270}
    \caption{[Fe/H] $\times$ $\log$ O/H + 12 relation for the
    galactic bulge (solid line) and solar neighbourhood (dotted line)
    from the [O/Fe] $\times$ [Fe/H] relations of Matteucci et al. 
    (1999). Also shown are: Four disk PN from Perinotto et al. (1999,
    squares with error bars), a disk PN from Pottasch \& Beintema (1999, 
    triangle with error bars), 6 bulge giants from McWilliam \& Rich (1994, 
    crosses), and a bulge star from Barbuy (1999, asterisk).}
    \label{fehoh}
    \end{figure*}

Chemical abundances of PN in the bulge of M31 have been recently
studied by Jacoby \& Ciardullo (\cite{jacoby}), who have included in
their analysis some results by Stasi\'nska et al. (\cite{stasinska}) and
Richer et al. (\cite{richer}). Fig.~\ref{pnoh} shows a comparison of this 
sample (42 PN, top panel) with the galactic bulge objects from Ratag et al. 
(\cite{ratag}) (103 PN, lower panel, thin line) and Cuisinier et al. 
(\cite{cuisinier}) (30 PN, lower panel, thick line). The oxygen abundance 
is given in the usual notation, $\epsilon({\rm O}) = \log ({\rm O/H}) + 12$.
It can be seen that all the O/H abundance distributions are similar, 
peaking around 8.7 dex, and showing  very few if any super metal rich 
objects with supersolar abundances. 

In order to compare the PN metallicity distribution with the stellar 
distributions, it is necessary to convert the measured nebular O/H 
abundances into the usual [Fe/H]  metallicities relative to the Sun. 
Direct measurements are of limited usefulness, for two main reasons. 
First, a very small number of planetary nebulae have measured iron lines, 
due to their weakness and the relatively large distances of the nebulae, 
so that the derived values are more uncertain than the usual O, N
or Ar abundances. Second, all available measurements indicate a strong 
depletion usually attributed to grain formation, so that the measured Fe 
abundances should be considered as lower limits to the total abundances 
at the times of formation of the PN progenitor stars. Both these aspects 
are illustrated in Fig.~\ref{fehoh}, where we plot the [Fe/H] abundances 
against oxygen for a group of four disk planetary nebulae from Perinotto 
et al. (\cite{perinotto}, squares with error bars) and one object by Pottasch 
\& Beintema (\cite{pottasch}, triangle with error bars). In all conversions 
we have used the solar iron abundance $\epsilon({\rm Fe})_\odot = 7.5$ (Anders 
\& Grevesse  \cite{anders}) and average O/H error bars.

A better way to convert O/H abundances into [Fe/H] metallicities is to 
use theoretical [O/Fe] $\times$ [Fe/H] relationships, such as those 
recently derived by Matteucci et al. (\cite{matteucci}) both for the 
solar neighbourhood and the galactic bulge. The corresponding relations 
are more suitably plotted in Fig.~\ref{fehoh} using 
$\epsilon({\rm O})_\odot = 8.9$ (Anders \& Grevesse \cite{anders}) both 
for the bulge (solid line) and the solar neighbourhood (dotted line).

    \begin{figure}
    \psfig{figure=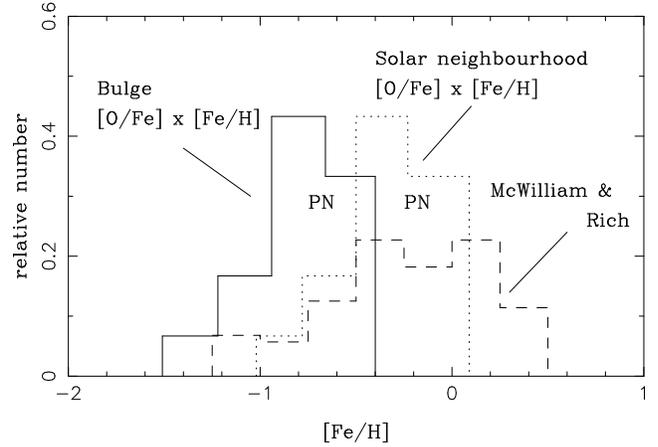,height=6.5 cm,angle=270}
    \caption[]{Metallicity distribution of bulge PN using the
    [O/Fe] $\times$ [Fe/H] relations from Matteucci et al. (1999)
    for the bulge (solid line) and solar neighbourhood (dotted line).
    The dashed histogram shows the distribution of K giants from 
    McWilliam \& Rich (1994).}
    \label{pnfeh}
    \end{figure}

Taking into account the relations shown in Fig.~\ref{fehoh}, the O/H 
distribution of bulge PN from Cuisinier et al. (\cite{cuisinier}) can 
be converted into a [Fe/H] distribution, as shown in Fig.~\ref{pnfeh}. 
Again the solid line corresponds to the [O/Fe] $\times$ [Fe/H] 
relationship for the bulge, and the dotted line was obtained using the 
solar neighbourhood relation shown in Fig.~\ref{fehoh}. As a comparison, 
the dashed histogram in Fig.~\ref{pnfeh} shows the metallicity distribution
of bulge K giant stars in Baade's Window by McWilliam \& Rich 
(\cite{mcwilliam}). It can be seen that the PN metallicity distribution 
looks similar to the K giant distribution if the {\it solar neighbourhood}
[O/Fe] $\times$ [Fe/H] relation is adopted, but when the {\it bulge} 
relation is taken into account the derived distribution is displaced 
towards lower metallicities by roughly 0.5 dex. The PN samples were
carefully selected not to include disk objects (see a detailed
discussion in Cuisinier et al. \cite{cuisinier}), and their progenitor
stars are clearly not massive enough to appreciably change their
initial O/H composition. Therefore, the bulge PN metallicity distribution 
is expected to be similar to that of the K giants. This conclusion is 
strenghtened by the earlier measurements of Sadler et al. (\cite{sadler}) 
for Mg and by the results recently presented by Feast 
(\cite{feast}), who has shown that the metallicity distribution of 
bulge Mira variables peaks around [Fe/H] $\simeq 0$, which is 
again higher by roughly 0.5 dex than the bulge PN shown by the solid 
histogram in Fig.~\ref{pnfeh}. Since both classes of objects
are basically the offspring of the evolution  of intermediate
to low mass stars, their metallicity distributions should also
be similar. These objects may be located within a certain distance 
from the galactic centre, but, as discussed by Frogel (\cite{frogel}), 
any gradient between Baade's Window and the Centre must be small, 
amounting up to a few tenths of a dex. 

\section{Discussion}

Several reasons can be considered in order to explain the discrepancy 
between the bulge PN distribution shown in Fig.~\ref{pnfeh} and the K giants 
of McWilliam \& Rich (\cite{mcwilliam}): (i) systematic errors in the 
O/H abundances of planetary nebulae, leading to mild to strong 
underestimates of this quantity; (ii) ON cycling in the PN progenitor 
stars, which would lead to a depleted O/H ratio; (iii) Statistical 
uncertainties due to the fact that the considered samples are relatively 
small and probably incomplete, and (iv) uncertainties in the adopted 
[O/Fe] $\times$ [Fe/H] relationships. Let us briefly consider each of 
these possibilities.

{\it First}, some recent work has raised the possibility that the O/H 
abundances of planetary nebulae may be underestimated, in view of the 
discrepancies between the forbidden line and recombination line values for 
a number of objects (Mathis \& Liu \cite{mathis}, Liu \& Danziger \cite{liu}). 
In principle, that would be applied to all PN samples considered in this 
work, namely, the bulge PN of Ratag et al. (\cite{ratag}), Cuisinier et al. 
(\cite{cuisinier}) and the objects in the bulge of M31 of Jacoby \& Ciardullo 
(\cite{jacoby}). There are no detailed published (under)abundance analyses 
for a large number of objects, but in order to be able to explain the 
discrepancy shown in Fig.~\ref{pnfeh}, the O/H abundances should have 
to be underestimated by about 0.5 dex, or a factor~3. This factor should 
also be applied to {\it disk} PN, as their abundances are derived using 
basically the same methods used for bulge PN.  However, Cuisinier et al. 
(\cite{cuisinier}) have shown that their metallicity distribution for 
bulge PN is similar to the corresponding  distribution of disk PN 
given by Maciel \& K\"oppen (\cite{mk94}). This is confirmed by the 
recent (although smaller) sample of disk PN used by Maciel \& Quireza 
(\cite{mq99}) to study radial abundance gradients in the galactic disk. 
This sample includes 128 disk PN, and  an application of the solar 
neighbourhood [O/Fe] $\times$ [Fe/H] relationship of Matteucci et al. 
(\cite{matteucci}) (dotted line in Fig.~\ref{fehoh}) produces a [Fe/H] 
distribution peaked at [Fe/H] $\simeq -0.3$. This is very similar to the 
recent G-dwarf metallicity distribution of the solar neighbourhood by 
Rocha-Pinto \& Maciel (\cite{rpm96}) within the usually adopted 
uncertainties in the PN abundances of up to 0.2 dex for oxygen. 
{\it Therefore,  if we were to apply a correction factor of about 0.5 dex
to the O/H abundances, the derived disk PN distribution would imply 
a very large number of extremely metal-rich objects, whose nature it would 
be very  difficult to explain. Moreover, the PN distribution would peak at 
about 9.2 dex, which means that most disk PN would  be more oxygen rich 
than  the Sun by about a factor 2}. Even considering that the populations 
of PN and G-dwarfs have somewhat different ages, no known age-metallicity 
relation  would be sufficient to explain such large overabundance 
(see for example the age metallicity relation by Twarog \cite{twarog} 
or the recent  work by Rocha-Pinto et al. \cite{rpsmf99} and references 
therein). As a conclusion, any  underestimate of the O/H abundances in PN 
would have to be much lower than 0.5 dex, and  could be well accommodated 
within an average uncertainty of 0.2 dex. Of course, this  does not exclude 
the possibility that some {\it individual} nebulae may have strongly 
underestimated abundances, but the {\it average} factor is probably much 
lower than 0.5 dex. Further work is needed, possibly including a 
sizable sample of disk PN with well measured forbidden-line and 
recombination-line abundances.

{\it Second}, the possibility of ON cycling has often been mentioned for 
planetary nebulae (see for example Maciel \cite{maciel92}), basically 
due to some anticorrelation in the N/O ratio compared with He/H for disk PN. 
However, this phenomenon is only expected to occur in Type I PN (cf. 
Peimbert \cite{peimbert}), which are formed mainly from the higher mass 
progenitor stars. These objects have an excess of He/H and/or N/O, and 
their oxygen abundances are in fact slightly lower than the most common 
Type~II objects (see for example Maciel \cite{maciel00}). However, the 
amount by which the O/H ratio is decreased is very small (roughly 0.1 dex), 
and Type I objects are explicitly excluded  from the disk sample of 
Maciel \& Quireza (\cite{mq99}). Regarding the bulge PN, almost all 
objects show no trace of He/H or N/O  enrichment, so that this possibility 
cannot be used to explain the discrepancy shown in Fig.~\ref{pnfeh}.

{\it Third}, statistical uncertainties are more difficult to analyze, since 
the considered samples are relatively small and may be affected by 
observational selection effects. However, all distributions can be 
understood in terms of general models for chemical evolution of the Galaxy, 
and the similarities of the metallicity distributions of different objects 
such as G-dwarfs and disk PN, or bulge PN, Mira variables and K giants 
fit rather nicely in the framework of galactic evolution, within the 
uncertainties of the derived abundances. Considering the bulge PN in 
particular, the similarity of the three different distributions shown in 
Fig.~\ref{pnoh} is striking, even though they reflect different samples 
using different techniques. Small differences such as the lack of
oxygen rich PN in the Cuisinier et al. (\cite{cuisinier}) sample may
be explained by the smaller size of this sample. However, we are interested 
in very broad aspects of these distributions, and not in the detailed 
behaviour at a given metallicity of range of metallicities, so that it is 
unlikely  that statistical uncertainties might produce the discrepancy  
shown in Fig.~\ref{pnfeh}.

{\it Fourth}, we are left with the possibility that the bulge [O/Fe] 
$\times$ [Fe/H] relation as given by Matteucci et al. (\cite{matteucci})
or, equivalently, the [Fe/H] $\times \log$ O/H relation given by the
solid line in Fig.~\ref{fehoh} might be responsible for all or most of the 
discrepancy in the metallicity distributions shown in Fig.~\ref{pnfeh}. 
Such relation assumes a faster evolution during the bulge formation,
so that at a given metallicity the [O/Fe] ratio is higher than in the 
solar neighbourhood. Although this is correct in principle, the present 
results suggest that  the amount of oxygen produced has been overestimated,
leading to an excess of O/H for a given metallicity. In fact, the earlier
models of Matteucci \& Brocato (\cite{mattbroc}) predict a lower [O/Fe]
enhancement, producing a better agreement with the present results. On the
other hand, there are some independent evidences that the theoretical 
[O/Fe] in the bulge may be overestimated. A recent analysis by Barbuy and 
collaborators for one star in the bulge globular cluster NGC 6528 
(see for example Barbuy \cite{barbuy})  shows that [O/Fe] $\simeq 0.35$ 
for a metallicity [Fe/H] $\simeq -0.6$, so that [O/H] $\simeq -0.25$ and 
$\log$ O/H + 12 $\simeq 8.65$. This object is also shown in 
Fig.~\ref{fehoh} (asterisk) and is closer to the dotted curve 
than to the bulge relation (solid line). Also, for {\it all} six bulge 
giants with measurable [OI] features in the McWilliam \& Rich 
(\cite{mcwilliam}) sample the [O/Fe] ratio is lower by $0.2-0.7$ dex 
than predicted by the [O/Fe] $\times$ [Fe/H] relationship of Matteucci 
et al. (\cite{matteucci}). These stars are also shown in Fig.~\ref{fehoh} 
(crosses), and are clearly located to the left of the bulge curve, even
allowing for some uncertainty in the O/H abundances. The 
average difference is about $0.5$ dex, which is just what is
needed to eliminate the offset of the bulge PN metallicity distribution
shown in Fig.~\ref{pnfeh}. Therefore, it can be concluded  that the 
theoretical [O/Fe] $\times$ [Fe/H] relation  for the bulge probably  
overestimates the oxygen enhancement relative to iron by 0.3 to 0.5 dex, 
at least for metallicities \hbox{[Fe/H] $\geq -1.5$ dex}.

\begin{acknowledgements}
       I am indebted to B. Barbuy, F. Matteucci, F. Cuisinier and R.D.D. 
       Costa, for some helpful discussions. This work was partially 
       supported by CNPq and FAPESP.
\end{acknowledgements}

   \end{document}